\documentclass[journal=nalefd,manuscript=letter]{achemso}
\setkeys{acs}{maxauthors = 0}      

\setkeys{acs}{articletitle = true} 
\usepackage[version=3]{mhchem} 
\usepackage[T1]{fontenc}       
\usepackage{SIunits}

\author{Aleandro Antidormi}
\affiliation[ICN2]{Catalan Institute of Nanoscience and
Nanotechnology (ICN2), CSIC and BIST,
Campus UAB, Bellaterra, 08193, Barcelona,
Spain}
\email{aleandro.antidormi@icn2.cat}
\phone{}
\author{Luciano Colombo }
\affiliation[Cagliari]{Dipartimento di Fisica, Universit\`{a} di Cagliari, Cittadella Universitaria, I-09042 Monserrato (Ca), Italy}
\author{Stephan Roche}
\affiliation[ICN2]{Catalan Institute of Nanoscience and
Nanotechnology (ICN2), CSIC and BIST,
Campus UAB, Bellaterra, 08193, Barcelona,
Spain}
\alsoaffiliation[ICREA]{ICREA Institucio Catalana de Recerca i Estudis Avancats,
08010 Barcelona, Spain}

\title{Thermal Transport in Amorphous Graphene with Varying Structural Quality}
\keywords{amorphous graphene, thermal transport, structural disorder, vibrational properties, thermal conductivity, 2D material }

\nocite{*}
\begin{tocentry}
\begin{center}
\includegraphics[scale=1]{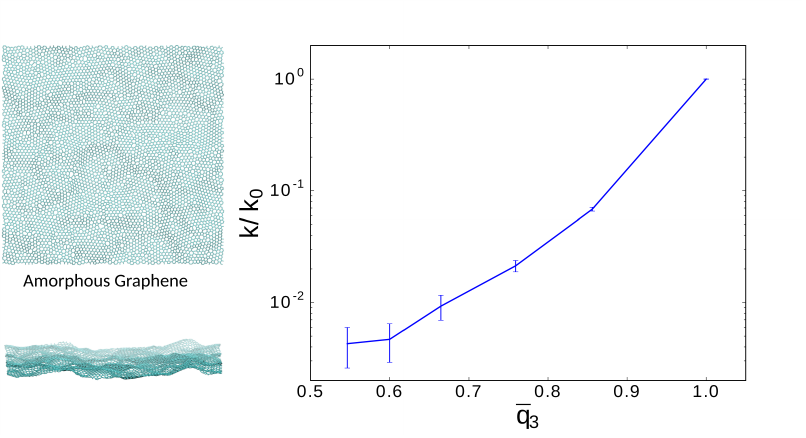}
\end{center}
\label{fig:toc}
\end{tocentry}

\begin{document}

\begin{abstract}
\normalsize
The synthesis of wafer-scale two-dimensional amorphous carbon monolayers has been recently demonstrated. 
This material presents useful properties when integrated as coating of metals, semiconductors or magnetic materials, such as enabling efficient atomic layer deposition and hence fostering the development of ultracompact technologies.
Here we propose a characterization of how the structural degree of amorphousness of such carbon membranes could be controlled by the crystal growth temperature. We also identify how energy is dissipated in this material by a systematic analysis of emerging vibrational modes whose localization increases with the loss of spatial symmetries, resulting in a tunable thermal conductivity varying by more than two orders of magnitude.
Our simulations provide some recipe to design most suitable "amorphous graphene" based on the target applications such as ultrathin heat spreaders, energy harvesters or insulating thermal barriers.
 \end{abstract}

\newpage


\section*{Introduction}

In recent years the fabrication, characterization and use of disordered forms of graphene-based materials has become a major field of concern, given the demonstrated capability for large scale production as well as the robustness of some physical properties unique to pristine graphene \cite{Ref1, Ref2, Ref3, doi:10.1021/acsnano.9b02621, PhysRevLett.106.105505}.
In particular, wafer-scale polycristalline graphene structures have been found to be essential for envisioning disruptive applications in sensing, photonics, electronics or spintronics \cite{doi:10.1021/acsnano.7b02474, Ref3, doi:10.1021/acsnano.6b08027, Ref4, doi:10.1021/acsnano.7b06800, Seifert_2015, Barrios_Vargas_2017}.
Other forms of disordered graphene (reduced graphene oxides), obtained by chemical exfoliation techniques, have been also found suitable for improving the performances of composite materials, with applications in energy and sensing\cite{doi:10.1021/acsnano.6b02391, Ref5, Ref6, Muchharla_2014, Donarelli_2015, Barrios_Vargas_2017}.
However, these forms of disordered materials are difficult to tailor and the quality of end products can greatly  vary even for a similar systhesis process \cite{Ref7, doi:10.1021/nn901689k}. \\
\indent In this context, the possibility to wafer-scale produce amorphous forms of carbon monolayers, structurally dominated by   sp$^2$ hybridization \cite{Jooe1601821,toh2020synthesis}, not only provides a new platform for exploring two-dimensional physics of strongly disordered materials, but also alternative forms of membranes with enhanced chemical reactivity which could serve as coating materials \cite{ozyilmaz2019two}.
Some theoretical models have been conceived either based on topological assembly rules or derived from molecular dynamics simulations \cite{doi:10.1002/pssb.200945581, PhysRevB.84.205414, ea86050b049f4b9a831ab75afbe07d5f, C8CP02545B}. If their electronic properties reveal a strong insulating behaviour \cite{PhysRevB.86.121408}, the variability of their vibrational and thermal properties \emph{versus} the degree of disorderedness remain to date unexplored. 
This is in contrast with other defective forms of graphene, like polycrystalline graphene and defect-engineered graphene nanoribbons, in which the role of the grain boundaries and the size of the grains in suppressing the thermal conductivity have been extensively investigated \cite{doi:10.1021/acs.nanolett.6b04936, doi:10.1021/acs.nanolett.7b01742, BAZRAFSHAN2018534, C3NR06388G, MORTAZAVI2013460}. \\
\indent As far as amorphous graphene (a-G) is concerned, some works have provided evidence that this material presents high elastic modulus ($\sim$500 GPa) and tensile strengths ($\sim$50 GPa) and a corresponding thermal conductivity as low as 15 W/mK \cite{MORTAZAVI2016318}. However, little (if any) physical information is available about the impact of the degree of disorder on (i) the spatial character of the microscopic heat carriers in a-G and (ii) on the vibrational and thermal properties of the material.  \\
\indent Here, using classical molecular dynamics we design large scale models of sp$^2$ carbon monolayers with a varying degree of structural irregularities, which quantify the deviation from an hexagonal lattice. We characterize the degree of disorder in real and reciprocal space and follow how vibrational properties evolve with increasing the loss of crystallinity.
We identify the class of vibrational modes emerging in such structures giving a complete picture of propagons, diffusons and locons. Finally, we connect their formation with the resulting thermal properties of those membranes through the application of a recently developed modal analysis of thermal conductivity.
Compared to the pristine graphene value, by tuning the crystalline order, the thermal conductivity is found to vary by more than two orders of magnitude, although remanining quite high compared to other amorphous materials: in amorphous silicon and amorphous carbon, for example, a suppression of more than three orders of magnitude is observed \cite{zhu}.

\section*{Structural and Vibrational Analysis}

In order to generate amorphous graphene samples, we followed a simulated \emph{quench-from-the-melt} method applied to a perfectly crystalline graphene plane (containing 10032 atoms) with square shape ($L= L_x = L_y  \approx 160 \angstrom$)\cite{ravinder2019evidence, VANHOANG201550, Kumar_2012}. 
The dimensions of the simulated systems have been chosen so as to meet three criteria, namely: (i) they are large enough to contain a reliable sampling of all possible bond orders and atomic coordinations likely present in an a-G sample, (ii) they allow for a converged estimation of thermal conductivity, and (iii) the required computational cost is not prohibitive.
All the simulations are performed within the classical molecular dynamics (MD) framework using the open source LAMMPS package \cite{plimpton}. The details of the preparation of the samples as well as the size-dependence of the results are given in the Methods section. \\
 \begin{figure*}[]
\begin{center}
\includegraphics[scale=0.5]{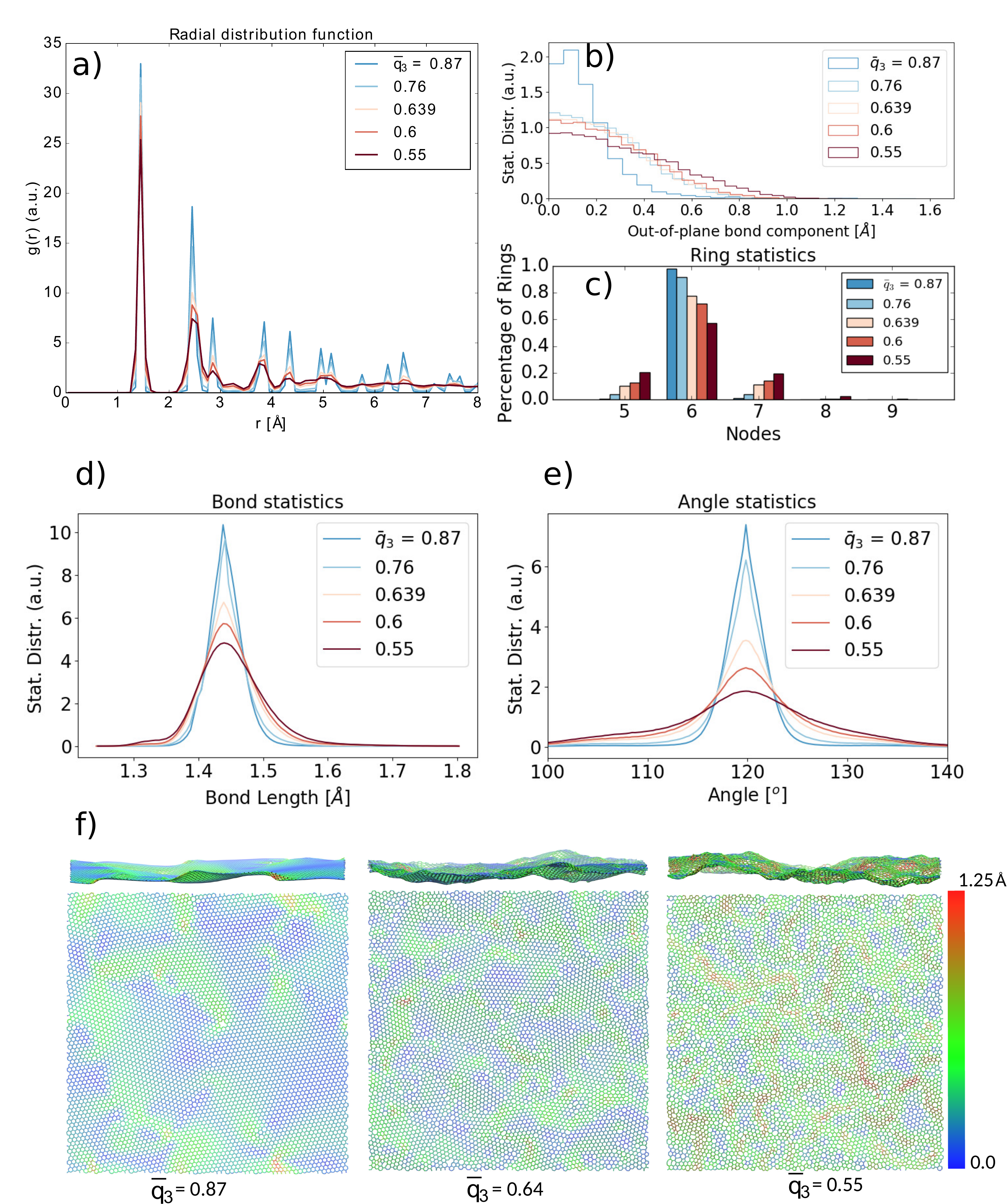}
\end{center}
\caption{a) Radial Distribution function $g(r)$ for different values of $\bar{q}_3$; b) statistical distribution of the out-of-plane component of the C-C bonds; c) Ring statistical distribution in a-G samples; d) Statistical distribution of the C-C bond length; e) Statistical distribution of bond angles; f) Structural models of three different samples of Am-G with the color map representing the out-of-plane component of the C-C bonds. }
\label{fig:structure}
\end{figure*}

\indent As a first result, we present the analysis of the structural properties of the a-G samples built in this work.
In particular, we analyze the radial distribution functions $g(r)$ (RDFs) of the samples \emph{versus} the degree of amorphousness (Fig.~\ref{fig:structure}(a)), quantified by means of the averaged triatic order parameter $\bar{q}_3$.  
This parameter (see the Methods for its definition) captures the deviation of the amorphous bonding network from the ideal sp$^2$-hybridized structure of pristine graphene by taking smaller (larger) values for more (less) amorphous samples, respectively. 

All the systems are characterized by the same average interatomic distance between nearest-neighbour atoms, as suggested by the same position of the first peak in the RDF found at $\sim$1.42$\angstrom$. In turn, a visible braodening affects the higher-order peaks, denoting a correspondigly wider distribution of distances between second- and third-nearest neighbours. The width of the peaks also increases with the degree of amorphousness as a consequence of the stronger structural disorder. 
Finally, for distances larger than 4$\angstrom$  and sufficiently amorphous samples, no peak can even be identified, implying a complete lack of long-range order in amorphous Graphene.  
The local character of a-G can also be appreciated from the statistical distribution of C-C bond lengths and angles (Fig.~\ref{fig:structure}(d-e)): their values are centered at 1.42 $\angstrom$ and 120$^o$, respectively, with a broadening which increases with disorder. These data are in actually good agreeement with similar characterizations of experimental samples\cite{toh2020synthesis, eder2014journey}.
 \\
\indent From the structural point of view, we also find that the a-G structures exhibit some static intrinsic ripples which compromise the planarity of the samples: more specifically, strongly wrinkled structures correspond to the more amorphous ones. This is clearly shown by the statistical distribution of the out-of-plane component of the C-C bonds, (see Fig. \ref{fig:structure}(b)): a larger tail in the gaussian distribution is observed upon increasing amorphousness. The degree of amorphousness is strongly related to the relative occurrence of 5-fold, 6-fold and 7-fold carbon rings: Fig.~\ref{fig:structure}(c) shows the percentage of $n$-fold rings in the structures, where small values of $\bar{q}_3$ are naturally linked to a larger number of non-hexagonal rings.  \\
\indent Further insight in the wrinkled character of the a-G sheet can be gained from the images of the their atomistic configurations (Fig.~\ref{fig:structure}(d)). In particular, we note that wrinkles and planarity deviations are localized in the structure where non-hexagonal rings are found (green regions in the figures). Such structural deformations, which mainly consist in a local curvature in the sheet, are energetically induced by the formation of the same topological defects. Analogous structural modifications have been observed also in carbon nanotubes and graphene with Stone-Wales defects \cite{Dettori_2012}.  \\

\begin{figure*}[]
\begin{center}
\includegraphics[scale=0.35]{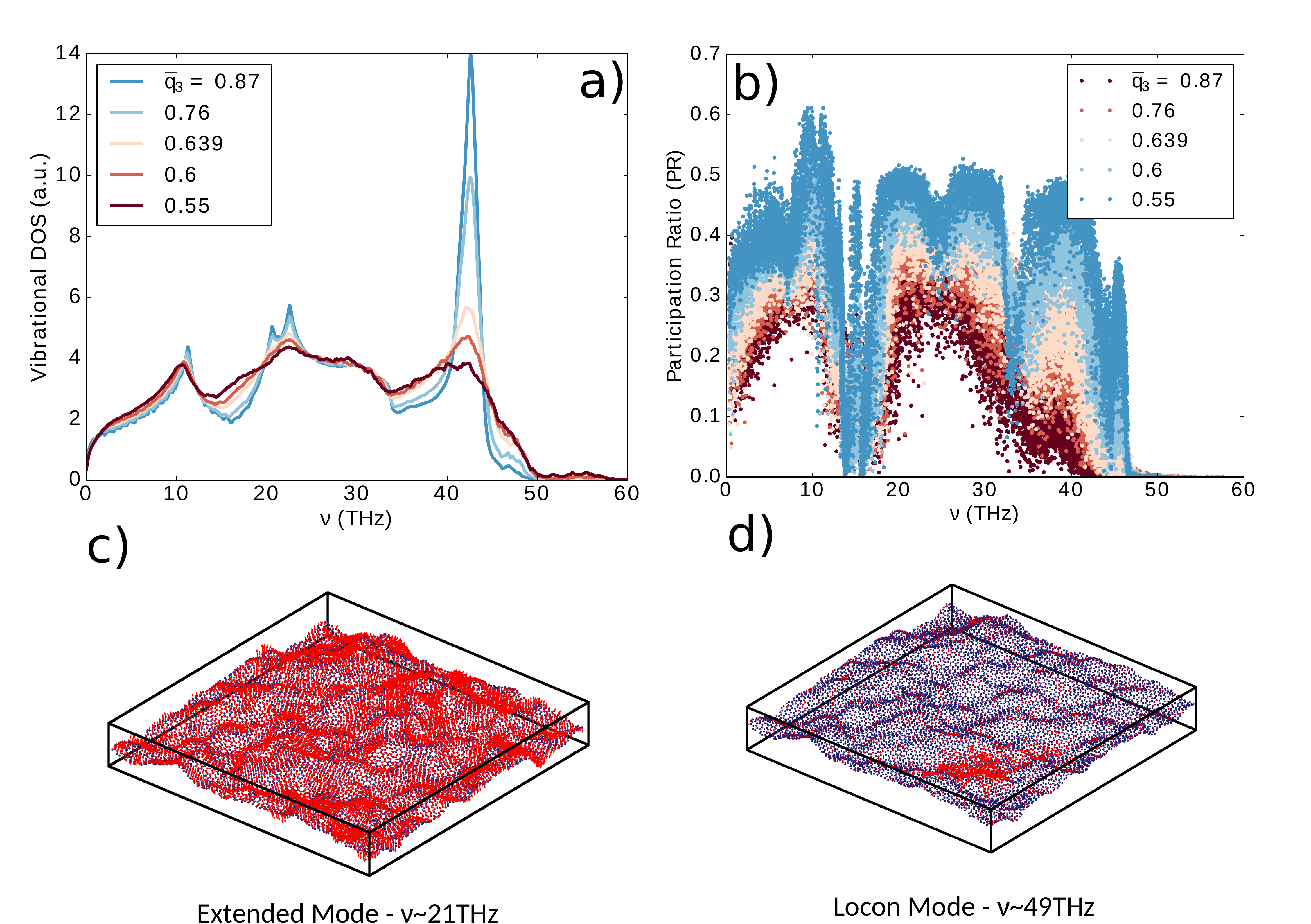}
\end{center}
\caption{a) The VDOS for a-G samples with different degree of amorphousness. b) Participation ratio (PR)of the eigenmodes \emph{vs.} frequency in a-G sasmples with different degree of amorphousness. c) and d) Atomic displacements for two eigenmodes in the sample with $\bar{q}_3=0.55$: a diffuson on the left and a locon on the right. The (properly scaled) displacements are depicted as red vectors superimposed on blue dots representing atoms. }
\label{fig:dos}
\end{figure*}

In Fig.~\ref{fig:dos}(a) we present the vibrational density of states (VDOS) of the five a-G systems with $ 0.55 \leq \bar{q}_3 \leq 0.87$. All the VDOSs share a similar behaviour, with peaks corresponding to different phonon bands. In low-disordered samples, the peaks converge to the corresponding features of the VDOS of crystalline graphene. The observed peak-resolved character of the vibrational density is, in turn, gradually lost as an effect of disorder: the peaks' broadening increases with amorphousness and eventually the different peaks merge into a \emph{continuum} of states for $\bar{q}_3=0.55$.\\

A more informative description of the vibrational modes is obtained by considering the Participation Ratio (PR) \cite{BELL1976215}, which offers a quantitative characterization of the spatial character of the eigenmodes. It is defined as \cite{allen, beltukov2}
\begin{equation}
PR_{\tt s} \equiv \frac{1}{N} \frac{\left(\sum_{i=1}^N e_{i, \tt s}^2\right) ^2}{\sum_{i=1}^N e_{i, \tt s}^4}
\end{equation}
where $e_{i, \tt s}$ is the $i$-th component of the $\tt s$-th eigenvector ($\tt s=1,..,3N$) and $N$ the number of atoms in the sample. The PR can take values in the interval [0,1] estimating the number of atoms participating to a given vibrational mode. In this respect, it allows to perform a classification of the modes on the basis of their spatial extension, distinguishing between extended modes, with $PR \sim 1$, and localized modes whose $PR$ approaches 0 in sufficiently large samples.
The PR of our systems is shown in Fig.~\ref{fig:dos}(b) as a function of the vibrational frequency.
 For the sake of clarity we do not show the PR of crystalline graphene which is equal to 1 in all the frequency spectrum.
In all the systems investigated, the participation ratio is smaller than 1, suggesting how a small structural perturbation is sufficient to alter the extended character of the phonon modes in an ideally crystalline sheet. \\
Fig.~\ref{fig:dos}(b) also clearly shows the impact of amorphousness on the spatial extension of the modes: a stronger disorder (smaller $\bar{q}_3$ ) leads to a generally smaller PR at all frequencies. 
This  result highlights the connection existing between the atomistic structure of the system and its vibrational properties. 
We remark that modes with particularly small participation ratio are found in correspondence of the band separation ranges among two adjacient graphene phonon bands, around $\nu\sim$15.0 THz and $\nu\sim$33.0 THz.
Importantly, a dramatic reduction of the participation ratio is observed for all the systems analyzed above $\nu\sim$47.0 THz. This frequency value corresponds to the so-called \emph{mobility edge}, which marks the separation between extended modes (also called \emph{extendons}) and localized modes, or \emph{locons}\citep{allen}.

\indent A more intuitive understanding of the difference between extendons and locons is gained by looking at their atomic displacements. To this aim, we show in Fig.~\ref{fig:dos}(c-d) the displacement field of a medium-frequency extendon and of high-frequency locon in the sample $\bar{q}_3=$0.55: scaled red arrows superimposed on each atom (in blue) denote the displacement vectors.
While in the extended mode, almost all the atoms participate to the motion, in the locon mode only a small subset of the atoms are sizeably vibrating around their equilibrium position.

We can further characterize the (extended) vibrational modes of a-G by defining  a wavevector, in close analogy to what is generally done for phonons in crystalline graphene. 
Based on this feature, Allen and Feldman \cite{allen} first proposed the distinction between \emph{propagons}, for which a wavevector can be meaningfully defined, and \emph{diffusons} which, on the contrary, do not allow such a definition.
This discrimination highlights a fundamental difference in the heat transport capabilities of these unalike vibrational modes: 
while propagons (which dominate the low-frequency part of the spectrum) spread out coherently along the sample for several interatomic distances before being scattered by disorder, diffusons (more commonly found at higher frequencies) scatter diffusively along the sample, as their name suggests.
In order to identify propagons and diffusons in our a-G systems, we compute the Fourier transform of the eigenvectors\citep{beltukov2}
\begin{equation}
F_{\eta}(\textbf{q}, \nu) = \sum_{\tt s=1}^{3N} \left| \sum_{i=1}^N  \theta_{i, \tt s, \eta}(\textbf{q}, \nu) e^{i \textbf{q \cdot R}_i}  \right| \delta(\nu-\nu_{\tt s})
\label{FT}
\end{equation}
where $\boldsymbol{R}_i$ is the position of the $i$-th atom whereas $\eta$ denotes the in-plane longitudinal (L), in-plane transverse (T) or out-of-plane flexural vibrational modes (Z) which characterize  the low-frequency vibrational spectrum of crystalline graphene.\\
 \begin{figure}[]
\begin{center}
\includegraphics[width=1.0\columnwidth]{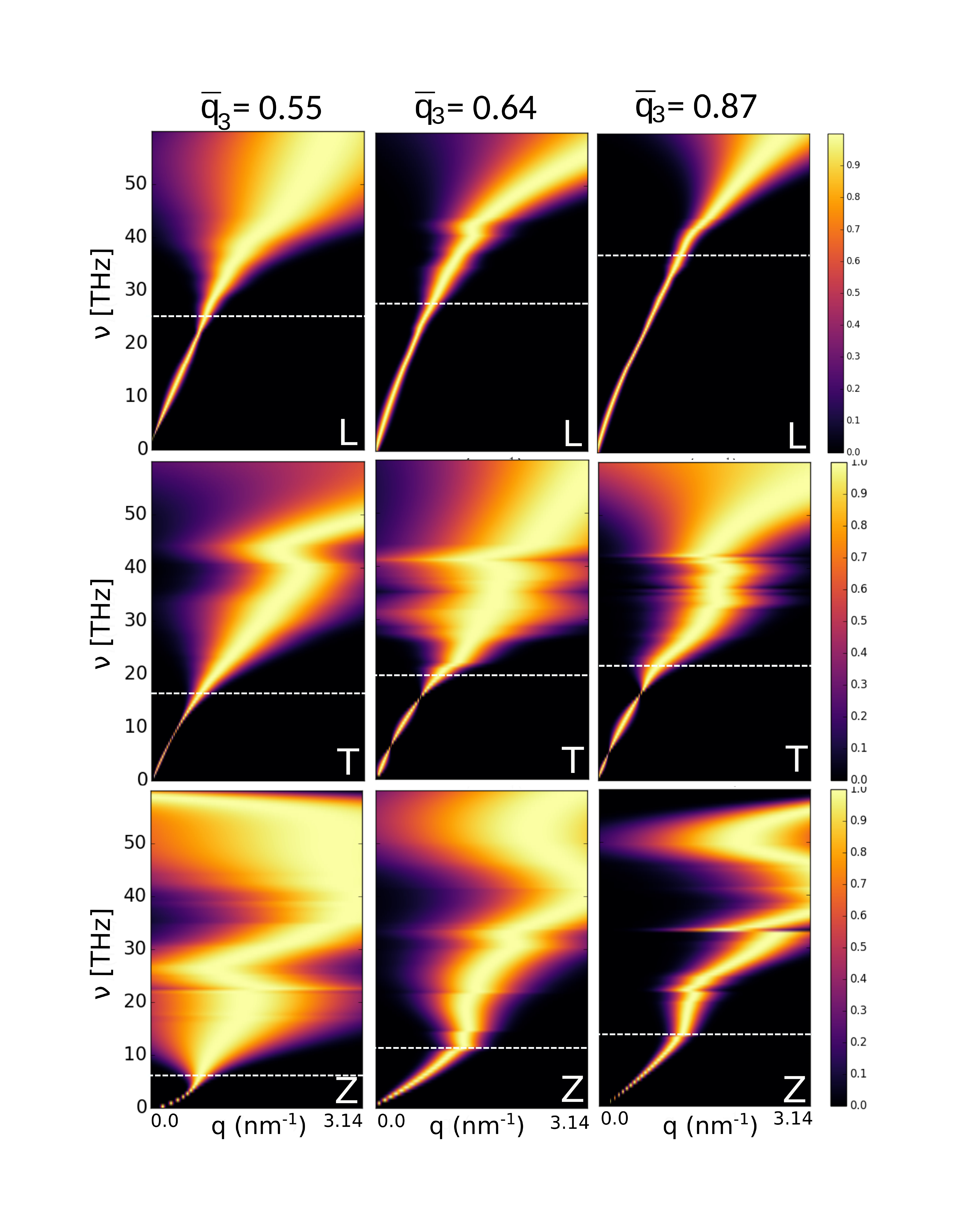}
\end{center}
\caption{Plot of $F(q, \nu)$ (right scale); $\nu^*=\nu^(q)$ dispersions (left scale) are visible connecting the maxima of  eq.~(\ref{FT}) for the different vibrational branches: $\eta$=L (top panel), $\eta$=T (middle panel) and $\eta$=Z (bottom panel) in the a-G samples with $\bar{q}_3$ of 0.55, 0.64 and 0.87,  respectively.}
\label{fig:k}
\end{figure}
In our systems,  $\theta_{L}(\textbf{q}, \nu) = \textbf{q}/|\textbf{q}| \cdot \textbf{e}_{i, \tt s} (\nu_{\tt s})$,
 $\theta_{T}(\textbf{q}, \nu) = \textbf{q}/|\textbf{q}| \times \textbf{e}_{i, \tt s, \parallel} (\nu_{\tt s})$ and $\theta_{Z}(\textbf{q}, \nu) = \textbf{q}/|\textbf{q}| \times \textbf{e}_{i, \tt s, \perp} (\nu_{\tt s})$, respectively. 
We sampled eq.~(\ref{FT}) on a grid of 20$\times$20 in a square region with side 8$\times$2$\pi$/L and finally averaged $F_{\eta}(\textbf{q}, \nu)$ over all directions of $\textbf{q}$ for each mode.
Fig.~\ref{fig:k} shows the plots of $F_{\eta}(q = |\textbf{q}|, \nu)$ with $\eta$=L (top panel), $\eta$=T (middle panel) and $\eta$=Z (bottom panel) for three different a-G samples with $\bar{q}_3$ of 0.55, 0.64 and 0.87,  respectively.
The function in eq.~(\ref{FT}) has been normalized with respect to its maximum value for each $\nu$ for a better graphical rendition.
\indent In the three cases shown, the contour of the maxima of $F(q, \nu)$ results in well-defined dispersion regions: a low-frequency region with a clear vibrational branch, and a high-frequency region where no clear relation between the wavector $q$ and the frequency $\nu$ is apparent. The frequency marking the separation between the two regimes is the Ioffe-Regel limit $\nu_{IR}$. Its position in the spectrum can be estimated,\citep{beltukov2, beltukov3} and is shown as white dashed lines in the plots of Fig.~\ref{fig:k}.
\\
\indent The effect of increasing the degree of amorphousness in graphene is clear: upon increase of disorder, low-frequency vibrational modes preserve a well-defined wavevector, whereas for higher frequency the dispersion curves become increasingly less defined, meaning that the definition of a wavevector becomes more and more questionable.
In particular, all the phonon branches of graphene undergo a lowering of the corresponding Ioffe-Regel frequency. Yet, an ordering relationship is still found to hold, namely $\nu_{IR}^Z < \nu_{IR}^T < \nu_{IR}^L$.

\subsection*{Thermal Transport Properties}

 \begin{figure}[h!]
\begin{center}
\includegraphics[width=1.0\columnwidth]{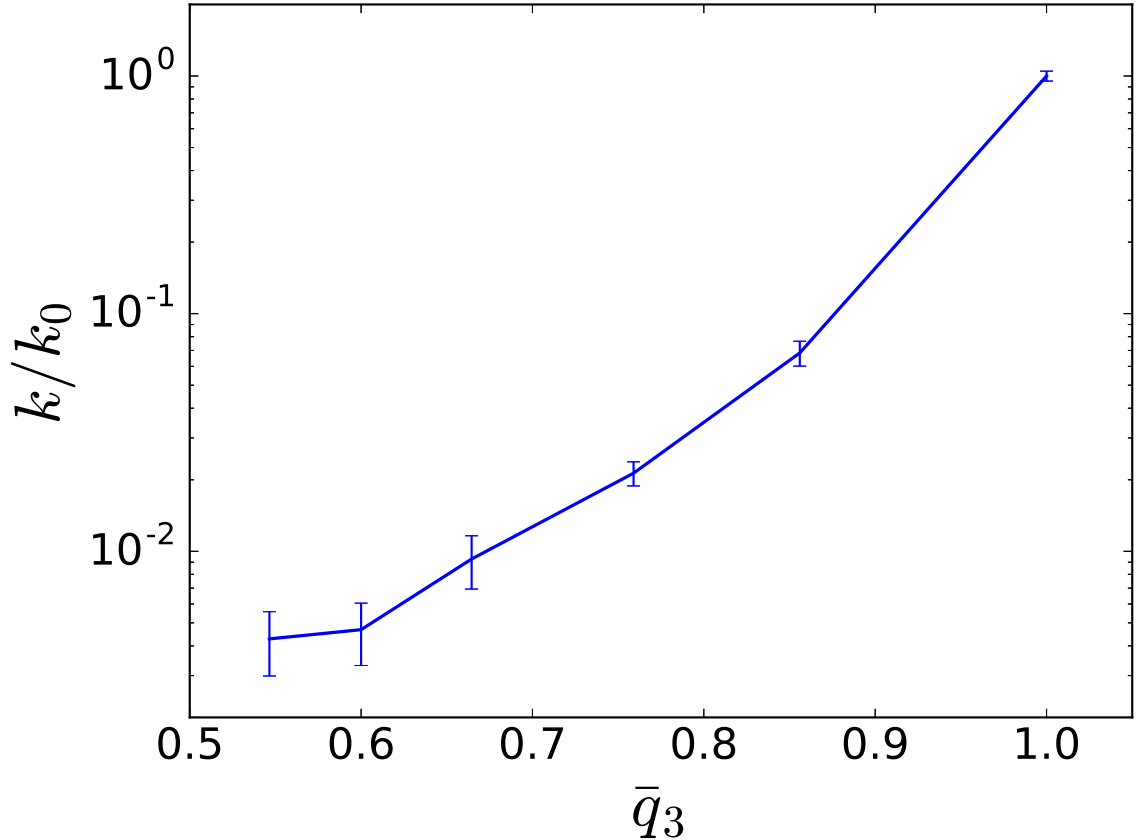}
\end{center}
\caption{Thermal conductivity of the a-G systems at T=300K \emph{vs.} $\bar{q}_3$.}
\label{fig:cond}
\end{figure}
We now present the results thermal transport properties of our a-G samples. All the calculations have been performed within the Green-Kubo Modal Analysis (GKMA)\cite{gkma} which allows to exploit the information on the vibrational eigenmodes already obtained to perform a modal decomposition of the thermal conductivity of the system and consequently assign a contribution to each class of modes. For further details on GKMA see the Methods section. \\
\indent We show in Fig.~\ref{fig:cond} the result of thermal conductivity calculation for a-G at 300K as a function of $\bar{q}_3$. 
The values are normalized with respect to the thermal conductivity calculated in a sample of crystalline graphene having the same dimensions of our a-G systems, namely $\kappa_0 = \kappa(L=160 \angstrom) \sim$ 2500 $\pm$ 120 W/(m $\cdot$ K). 
This value is in really good agreement with previous estimations, which range in the interval 2600-3050 W/(m $\cdot$K) \cite{xu2014length, MORTAZAVI20151, Bagri, MORTAZAVI2016318, PhysRevB.95.144309, doi:10.1021/nn200114p, doi:10.1021/nl0731872, ghosh2010dimensional, PhysRevB.92.094301, PhysRevB.81.205441, doi:10.1021/nl102923q} and have been obtained with a similar computational setup (see the Methods). 

\begin{figure}[]
\begin{center}
\includegraphics[width=1.0\columnwidth]{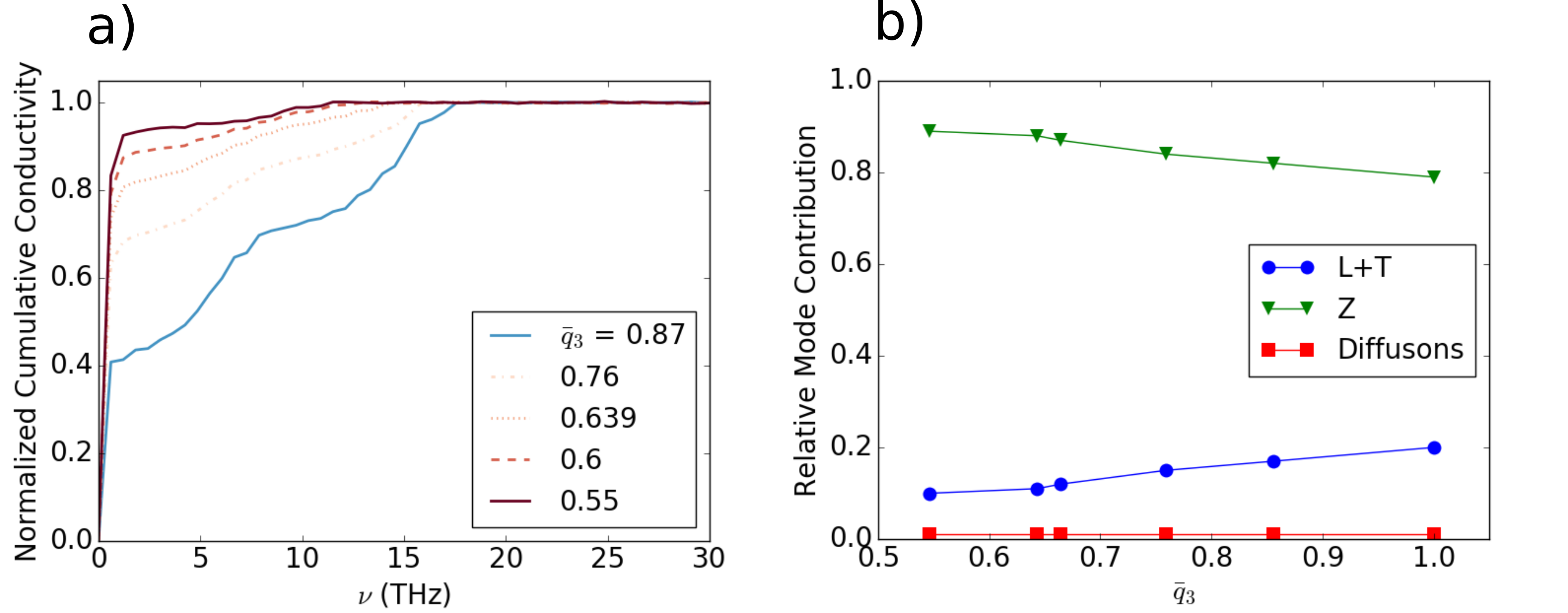}
\end{center}
\caption{a) Normalized thermal conductivity accumulation function of the a-G systems at T=300K. b) Relative contributions of the different classes of vibrational modes to the thermal conductivity in the a-G systems investigated.}
\label{fig:spectrum}
\end{figure}

The main result of Fig.~\ref{fig:cond}  is the strong increase of $\kappa$ as a function of the triatic order, showing how amorphousness dramatically inhibits thermal transport: a ratio of more than two orders of magnitude is, indeed, observed between the thermal conductivity in crystalline graphene and in the most amorphous sample. \\ 
 \indent To gain insight on the mechanisms underlying heat transport in a-G, in Fig.~\ref{fig:spectrum}-a we also show the  accumulation function of the thermal conductivity $\kappa$ of our systems, properly normalized to the pristine graphene value $\kappa_0$: this quantity represents the cumulative thermal conductivity of the vibrational modes up to a given frequency. 
The almost monotonical increase of the integral with frequency allows to assign a specific contribution to all different vibrational modes. 
In particular, we  observe how low-frequency modes are responsible for the largest contribution to the overall thermal conductivity in all the systems studied. Specifically, modes with frequencies smaller than 1THz contribute from 40\% of the total $\kappa$ in the sample with $\bar{q}_3=0.83$ to 93\% in the $\bar{q}_3=0.55$ system. 
We recognize that this is mainly due to the fact that low-frequency modes are more populated at finite temperature with respect to high-frequency modes and low-frequency modes are acknowledged to be more effective in carrying heat along the sample due to their propagon character. \\
\indent We also remark that for all the considered systems a value of frequency can be found in correspondence of which the cumulative conductivity saturates to unity. Such value of frequency strongly depends on the amorphousness and is large when $q_3$ is large (\emph{i.e.} less amorphous samples). 
This points to the non-negligible weight associated to high frequency modes in less amorphous a-G samples. 
It is also important to stress that the modes which contribute to the largest extent to $\kappa$ are found to lie below the out-of-plane flexural Ioffe Regel limit, which we calculated in the previous section. Inded, the dominant contribution to $\kappa$ is due to flexural modes (propagons whose atomic motion is mostly out of the graphene plane). \\
\indent In order to prove this assertion, we focus on the relative contribution of the different classes of modes to the thermal conductivity. Fig.~\ref{fig:spectrum}-b shows the decomposition of the overall thermal conductivity into the contributions of in-plane modes, flexural (Z) modes and diffusons, respectively.
Flexural modes are responsible for the largest part of the thermal conductivity in crystalline graphene (in agreement with other studies) and in all the amorphous systems considered in this work: their contribution increases with disorder ranging from 90\% in the most amorphous system to 80\% in the most crystalline one. 
Besides, the contribution of transverse and longitudinal modes is much weaker and decreases with the degree of amorphousness. In all the cases, the diffusons are generally less effective in carrying heat flux with respect to propagons and their impact is of the order of 1\% of the total thermal conductivity. 
Such finding helps clarifying the basic mechanism underlying the reduction of thermal conductivity in a-G with respect to crystalline graphene. The presence of disorder in the crystalline lattice determines an essential modification of the character of the low-frequency flexural modes (which are responsible of the largest part of thermal transport) by transforming their propagon character to diffusive one. The overall $\kappa$ is consequently reduced due to the less effective nature of diffusons in carrying heat. \\

\section{Conclusions}

\indent To summarize, in this work we study, within a classical molecular dynamics framework, the vibrational and thermal properties of amorphous graphene. By performing a detailed analysis of the vibrational modes of a-G samples, we analyze the impact of the degree of amorphousness showing how the progressive loss of crystallinity of the samples leads, on one side, to structural deviations from the planarity and, on the other, to a progressive transformation of the vibrational modes responsible for heat transport. 
Specifically, the transformation of the modes from propagons to diffusons \emph{via} a lowering of the corresponding Ioffe-Regel limits, ultimately leads to a reduction of thermal conductivity of a-G up to more than two orders of magnitude with respect to its crystalline counterpart.
These findings motivate the search of experimental strategies to control the degree of amorphousness in a-G in order to design a-G samples with improved performances depending on the specific applications.

\section*{Methods}

\subsection*{Sample Generation}

\begin{figure}[]
\begin{center}
\includegraphics[scale=0.7]{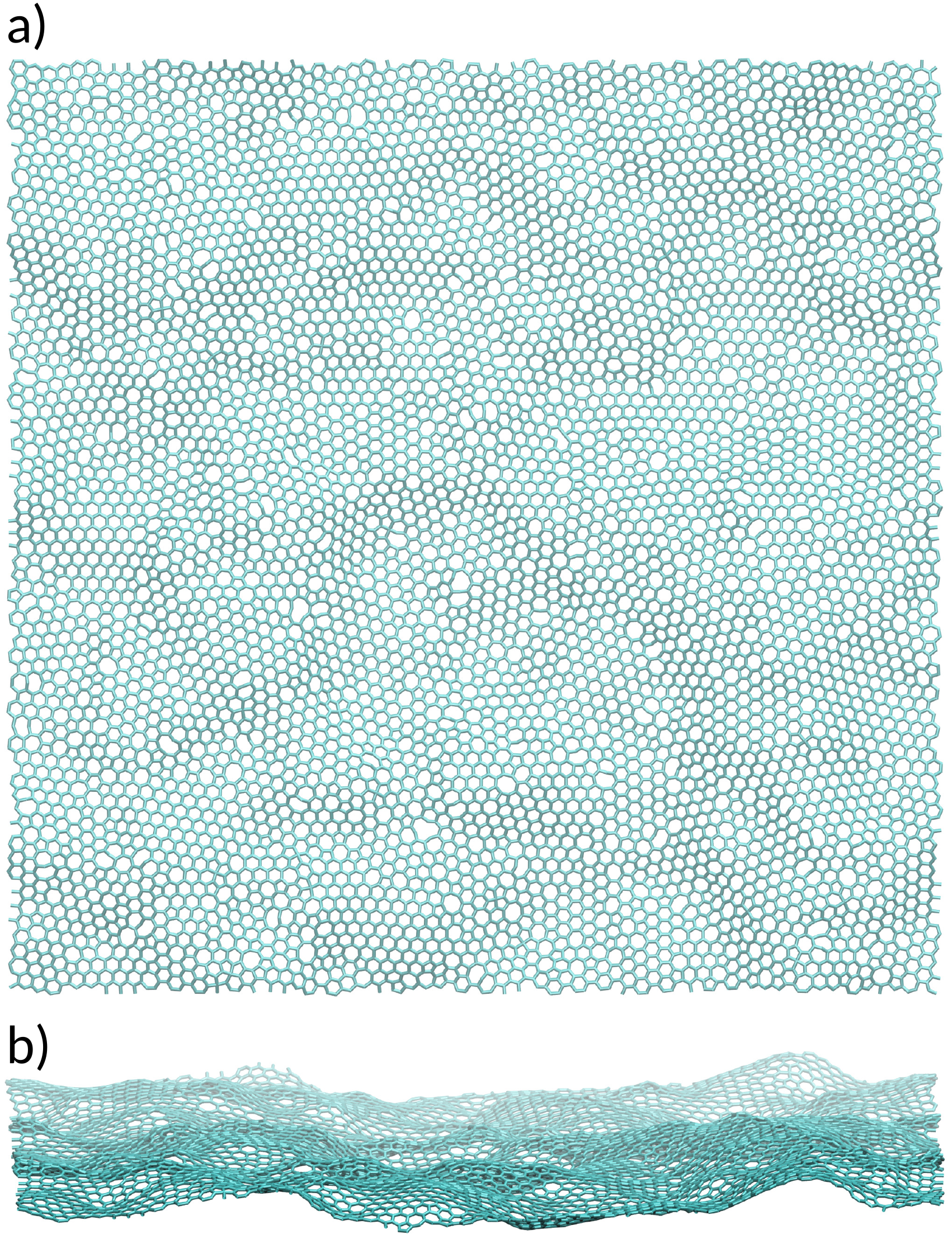}
\end{center}
\caption{Structure of the a-G sample with $\bar{q}_3=$ 0.639. Panel a: top view; panel b: cross-section.}
\label{fig:example}
\end{figure}

A multi-step procedure has been employed to generate samples with different degrees of amorphousness: first, the crystalline system is subjected to a 10 ns run (timestep 0.25fs) at constant temperature T=300K using a Nos\'{e}-Hover thermostat; then, its temperature is gradually increased in the constant-volume, constant-temperature (NVT) ensemble with a rate of 50 K/ps up to  10000 K at which the system is found in a two-dimensional liquid melt state. Next, the system is equilibrated at this temperature for 1 ns. All these simulations are performed by constraining the atomic motion to the original 2D plane.
The temperature of the system is then gradually reduced to 300 K with different cooling rates in the NVT ensemble. 
Depending on the cooling rate applied, amorphous systems are obtained differing in the amount of disorder or, equivalently, in the degree of amorphousness. 
In this work we built five different amorphous systems applying cooling rates values in the range [50K/ps, 1000K/ps].
The structures thus obtained are then relaxed at 300K for 1 ns and eventually equilibrated for 1 ns at the same temperature with the atoms left free to move also in the out-of-plane direction. 
A representative sample of a-G investigated in this work is shown in Fig.\ref{fig:example}.

Periodic boundary conditions have been applied along the in-plane directions and the evolution of the system is performed by ageing the atomic trajectories with the velocity-Verlet algorithm.
In all our simulations, we adopted the optimized Tersoff potential \cite{PhysRevB.81.205441} to describe the atomic interactions. This choice is supported by several studies assessing the reliability of this force-field \cite{SI2017450, khan2015equilibrium} to describe the thermal transport properties of various carbon forms, \emph{e.g.}, fullerene molecules, carbynes, cluster-assembled nanogranular Carbon films, polycrystalline graphene and, finally, Defective Graphene \cite{doi:10.1021/acs.nanolett.6b04936, doi:10.1021/acs.nanolett.7b01742, BAZRAFSHAN2018534, PhysRevB.92.094301}.\\

\begin{figure}[]
\begin{center}
\includegraphics[scale=0.7]{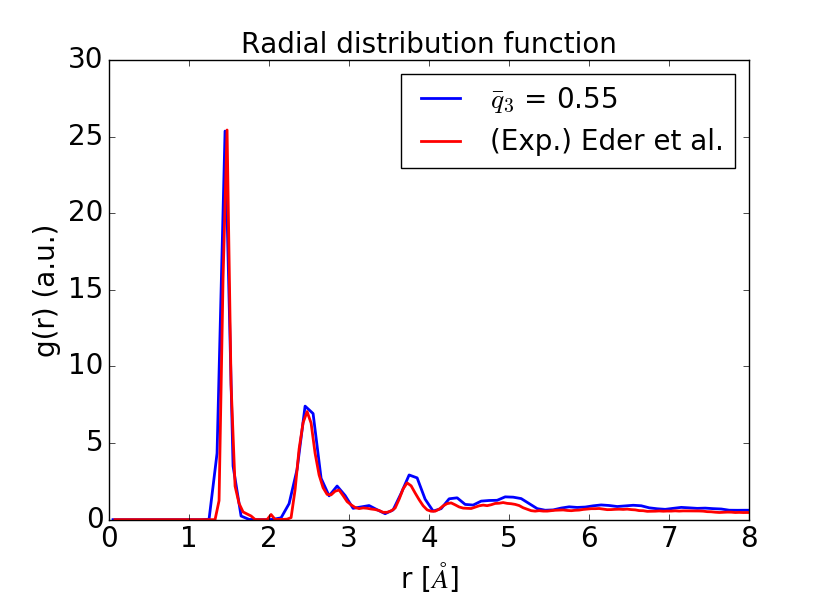}
\end{center}
\caption{Radial distribution functions of a-G obtained with a cooling rate of 1000K/ps (blue) and of the irradiated graphene sheet with 19.7\% of density deficit experimentally fabricated in \cite{eder2014journey} (red)}
\label{fig:comp}
\end{figure}

In order to assess the accuracy of the atomistic a-G samples and the reliability of the protocol employed, which could lead to unrealistic structural features upon high cooling rates, we compared the structure obtained with the highest cooling rate (1000K/ps) with the systems of irradiated graphene experimentally fabricated from Eder et al. in Ref.\cite{eder2014journey}. 
Fig.~\ref{fig:comp} shows the radial distribution function of our sample and of the irradiated graphene sheet with 19.7\% density deficit. The latter corresponds to the most disordered system discussed in Ref.\cite{eder2014journey}. Extremely good agreement between the two curves can be observed with respect both to the position of the peaks and their relative amplitude.
Both samples show a complete lack of structural order for distances larger than 6 $\angstrom$.

\subsection*{Structural and Vibrational properties}

\begin{figure}[]
\begin{center}
\includegraphics[scale=0.5]{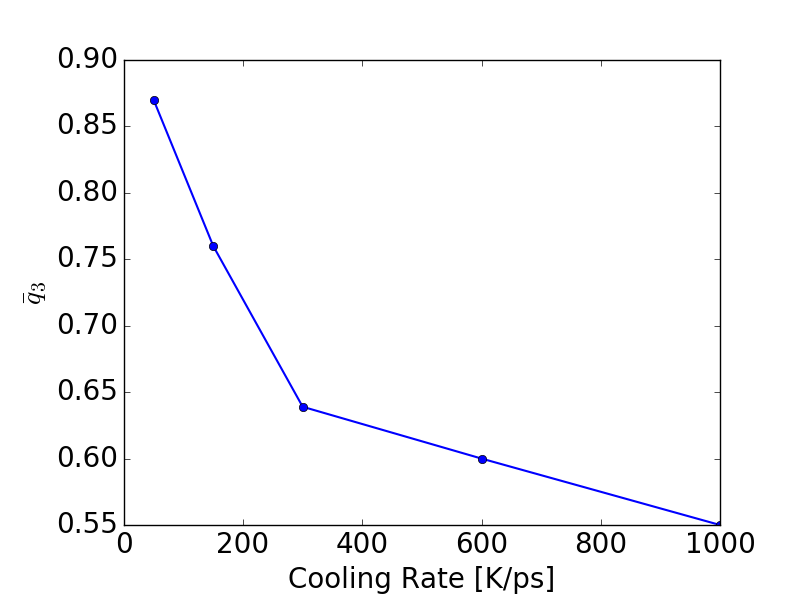}
\end{center}
\caption{Value of the triatic order $\bar{q}_3$ as a function of the cooling rate employed in the a-G sample generation.}
\label{fig:q3_vs_rate}
\end{figure}

In order to quantitatively characterize the degree of amorphousness of our systems, we compute the triatic order parameter $q_3$ \cite{PhysRevB.28.784, PhysRevLett.99.055702}, which effectively captures the deviation of the amorphous bonding network from the ideal sp$^2$-hybridized structure of pristine graphene. 
To this aim, we define the vector $q_{lm}(i)$ of atom $i$ as 
\begin{equation}
q_{lm}(i) = \frac{1}{N_n(i)} \sum_{j=1}^{N_n(i)} Y_{lm}(\boldsymbol{r}_{ij})
\end{equation}
Here, $N_n(i)$ is the number of nearest neighbors of atom $i$ - defined as the atoms found within a spherical volume with radius $r=2.3 \angstrom$ (corresponding to the second maximum in the radial distribution function),
 $l$  is an integer parameter with $m$ such that $-l \leq m \leq +l$. The functions $Y_{lm}(\boldsymbol{r}_{ij})$ are the spherical harmonics and $\boldsymbol{r}_{ij}$ is the vector from atom $i$ to atom $j$. Using $l=3$, we then define the triatic order for atom $i$ as 
\begin{equation}
q_3(i) = \sqrt{ \frac{4\pi}{2l+1} \sum_{m=-3}^{m=3} |q_{3m}(i)|^2}
\end{equation}
Finally, averaging over the N atoms in the system we obtain $\bar{q}_3$, which can take values in the range [0,1], with $\bar{q}_3=1$ for the perfectly crystalline graphene plane. 
Fig.~\ref{fig:q3_vs_rate} shows the value of $\bar{q}_3$ for our a-G systems as a function of the cooling rate employed. 
As it can be seen, slower cooling yields more ordered networks (larger $\bar{q}_3$); hence, changing the cooling rate allows us to tailor the degree of amorphousness in the structures and to probe its influence on properties.

We performed the analysis of the vibrational properties of our systems by calculating and diagonalizing their dynamical matrix, which is calculated as (we use Latin and Greek letters to denote atoms and Cartesian axes, respectively)
\begin{equation}
D_{i \alpha, j \beta} = \frac{1}{\sqrt{M_i M_j}}\frac{\partial F_{i\alpha}}{\partial r_{j\beta}}
\label{matrix}
\end{equation}
where $M_i$ is the mass of atom $i$ and $F_{i\alpha}$ is the projection of the total force acting on atom $i$ along the $\alpha$-direction. The derivative on the right of eq. \ref{matrix} represents the change in the $\alpha$-component of $F_{i}$ due to an infinitesimal displacement of the $j$-th atom along direction $\beta$ with respect to its equilibrium position. 
Its numerical evaluation has been carried out using a finite difference approximation with an atomic displacement of $5\cdot 10^{-4} \angstrom$. 
The resulting dynamical matrix has rank $(3N)^2$, with $N$ the number of atoms in the system, which in our case is of the order $\sim 10^5$. The numerical implementation of the diagonalization required the use of the SLEPc library \cite{hernandez2005slepc, petsc-web-page, petsc-user-ref, petsc-efficient} to get the 3N eigenvectors $\textbf{e}_{\tt s}$ and eigenvalues $\omega^2_{\tt s}$ ($\texttt{s}=1,..., 3N$). 

\subsection*{Thermal transport properties}

The information on the vibrational eigenmodes allows to determine the thermal conductivity of our samples and compute the relative contributions of the different classes of modes to the overall conductivity $\kappa$. We performed the modal decomposition of $\kappa$ within the formalism of the so-called  Green-Kubo Modal Analysis (GKMA)\cite{gkma}, already successfully exploited to study the vibrational properties of crystalline and amorphous silicon \cite{gkma}, porous silicon\cite{PhysRevMaterials.2.056001}, amorphous carbon \cite{doi:10.1063/1.4948605} and polymers\cite{PhysRevMaterials.4.035401}.
We refer the reader to Ref. \citep{gkma} for the details of the methodology.
Combining the outcome of the dynamical matrix diagonalization and long MD runs, GKMA permits to write the total thermal conductivity $\kappa$ as 
\begin{equation}
 \kappa = \sum_{\texttt{s}=1}^{3N} \kappa(\texttt{s}) =    \sum_{\texttt{s}=1}^{3N} \frac{\Omega}{3k_BT^2} \int_0^\infty \left\langle \textbf{Q}_{\tt s}(t) \cdot \textbf{Q}(0) \right\rangle dt 
\label{TC3}
\end{equation}
Here, $\Omega$ is the volume of the simulation cell, $k_B$ is the Boltzmann constant and $T$ the temperature. In eq. \ref{TC3}, the total thermal conductivity is expressed as the sum of single-mode contributions $\kappa(\texttt{s})$ given by the time-correlations between the single-mode heat flux operators $\textbf{Q}_{\tt s}(t)$ \citep{PhysRev.132.168} and the total heat flux operator. The single-mode heat flux operator is a function of the atomic positions and velocities\citep{PhysRev.132.168}:  
\begin{equation} 
\textbf{Q}_{\tt s}(t) = \frac{1}{\Omega} \sum_{i=1}^N \left[ E_i \dot{\textbf{x}}_i(\texttt{s},t) + \sum_{k=1}^N (\textbf{F}_{ik} \cdot \dot{\textbf{x}}_i( \texttt{s},t)) \textbf{r}_{ik}\right]
\end{equation}
where $E_i$ is the sum of the potential and kinetic energy of atom $i$, $\textbf{r}_{ik}$ is the distance vector between atoms $i$ and $k$, and $\textbf{F}_{ik}$ is the force acting on atom $i$ due to the interaction with atom $k$.
The information on the eigenvector is contained in the quantity $\dot{\textbf{x}}_i(\tt s,t) = (\textbf{v}_i(t) \cdot \textbf{e}_{i, \texttt{s}}) \textbf{e}_{i, \texttt{s}}$, obtained projecting the atomic velocities onto the $\texttt{s}$-th eigenvector during a constant-energy MD trajectory.

\indent In order to take into account the error that a classical calculation generally introduces in the value of the specific heat in materials below their Debye temperature, a quantum heat capacity correction has been introduced in eq.~(\ref{TC3}): the classical Dulong-Petit specific heat  $c_{DP}(\omega)=k_B/\Omega$ has been replaced by its quantum counterpart  $c_q(\omega) = \frac{k_B x^2 \exp(x)}{\Omega (\exp(x)-1)^2}$, where  $x=h\omega/(k_BT)$ \cite{ziman, doi:10.1063/1.4948605}.
No quantum correction to phonon lifetimes has been applied, as it would be in the case of a perfectly crystalline system whose phonon populations are incorrectly sampled by the classical trajectories generated by MD. Instead, in amorphous systems as the presently investigated ones, it has been found\cite{doi:10.1098/rspa.1951.0147, doi:10.1063/1.4948605} that the lifetimes of the microscopic heat carriers are mainly limited by the short-range structural disorder rather than by intrinsic anharmonicity. 
This feature is common to other situations discussed in the literature (and similarly treated as we do in our investigation) where the phonon scattering by lattice defects, such as boundary \cite{doi:10.1063/1.119402}, impurity \cite{PhysRev.134.A471} or structural scattering \cite{doi:10.1098/rspa.1951.0147, doi:10.1063/1.4948605}, dictates phonon relaxation times. 
In such situations (\emph{i.e.} whenever anharmonic scattering processes are not the primary mechanism limiting phonon lifetimes), it is expected that the error performed by classical MD in evaluating at low temperatures (\emph{i.e.} below the Debye temperature) the populations of the various phonon modes (\emph{i.e.} the true quantum feature underlying their scattering) could become negligible. 
Finally we remark that no quantum correction has been applied to the results of crystalline graphene, since the classical estimation of its thermal conductivity at room temperature has been shown to be only 10\% smaller than the corresponding quantum value\cite{doi:10.1063/1.3665226}. This results from the compensation of two errors arising from a classical calculation, i.e. shorter phonon lifetimes but larger heat specific heat with respect to their quantum counterparts.

\begin{figure}[]
\begin{center}
\includegraphics[scale=0.5]{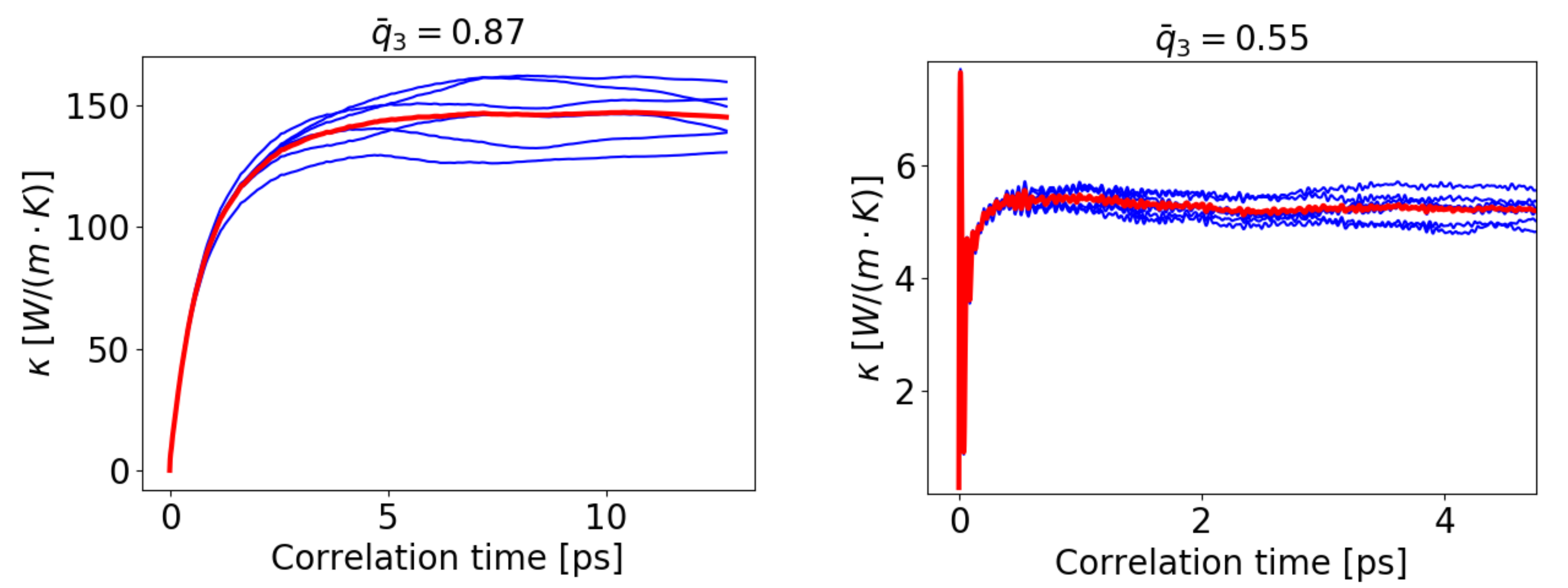}
\end{center}
\caption{Calculated Green-Kubo thermal conductivity (with no quantum corrections) as a function of correlation time for a-G samples with $\bar{q}_3=0.55$ and $0.87$ at 300 K. Blue lines represent the results of independent simulations while the red thick lines are the ensemble average.}
\label{fig:correlation}
\end{figure}

To compute the GK thermal conductivity of a-G ( eq.~(\ref{TC3})) we evolved the systems for a simulation time as long as 10ns, after a preliminary equilibration at T=300K for 2 ns with a 0.25 fs timestep. Six independent trajectories with different sets of initial atomic velocities have been analyzed for each sample. This choice has proved to provide a good statistical average in all the investigated samples, independently of the amount of disorder. Indeed, the maximum correlation time of the heat flux in the least disordered systems is smaller than 20ps. This is shown in Fig.~\ref{fig:correlation} where the thermal conductivities calculated for two samples of a-G as a function of the correlation time are depicted.

\subsection*{Validation steps}

\begin{figure}[]
\begin{center}
\includegraphics[scale=0.6]{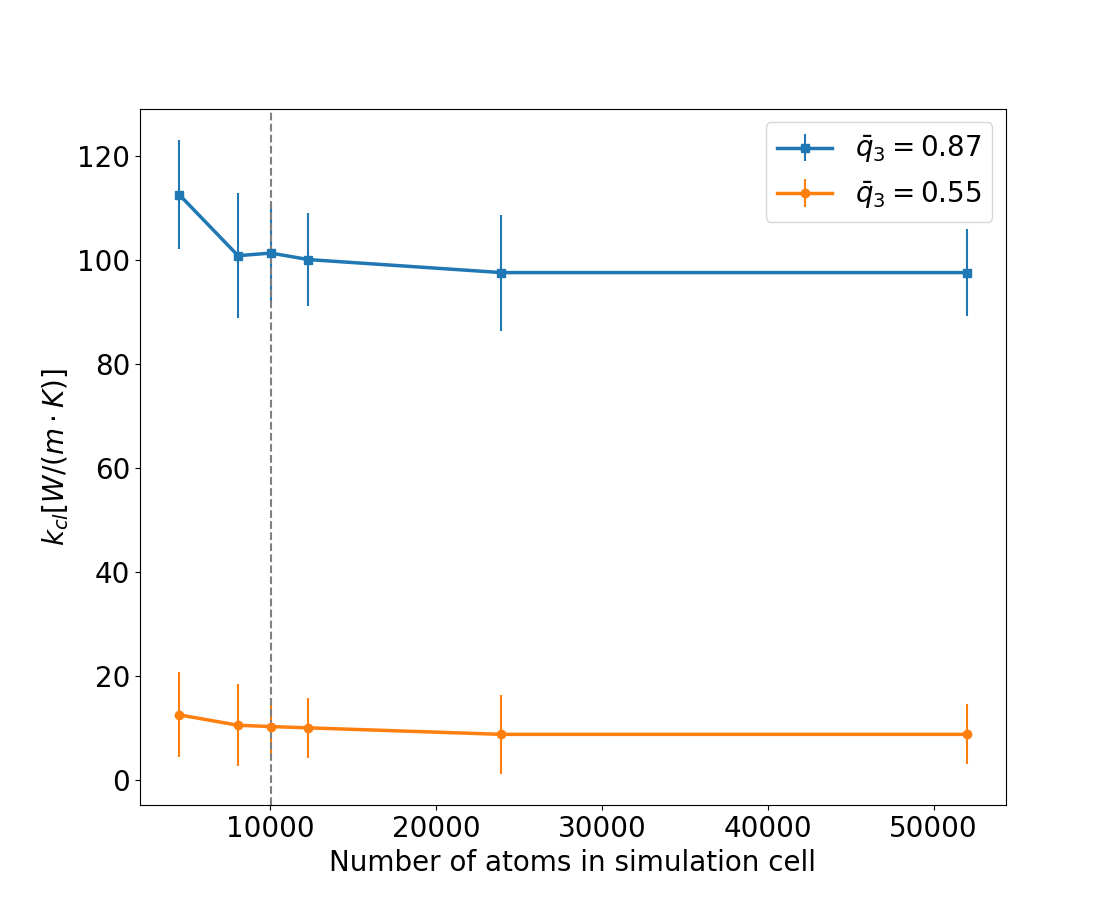}
\end{center}
\caption{(colSuor online) Green-Kubo thermal conductivity (with no quantum corrections) for a set of a-G samples with $\bar{q}_3=0.55$ and $0.87$ as a function of the system size. The dashed vertical line denotes the size of the systems investigated in this work.}
\label{fig:size-analysis}
\end{figure}

As a preliminary step, we investigated the size-dependence of the thermal properties of a-G and verified that dimension of the samples here analyzed is large enough to guarantee that all the properties calculated are well-converged with respect to the system size. 
In order to show this, we focus on the room-temperature thermal conductivity by the Green-Kubo (GK) method (without quantum corrections) for two systems with unalike value of triatic order. In Fig.~\ref{fig:size-analysis}, the values of thermal conductivity are shown as a function of the number of atoms in the simulation cell.
Each point of the curve is the average over three different systems generated independently and characterized by the same value of $\bar{q}_3$ (within a 0.02 uncertainty). The conductivity of each system is computed using six trajectories as long as 10ns. In both case studies, the size of the systems employed in our work (marked by a vertical dashed line) lies well within the range of convergence. 
Eventually, the selected system size to perform our investigation was set to $\sim$10000 atoms, so as to keep the computational burned at a reasonable level. \\

We also performed the calculation of the Green-Kubo thermal conductivity of crystalline graphene as benchmark for the adopted force-field. The corresponding calculated value $\kappa_0$ is used in this work as a reference for the conductivity of a-G samples. 
Specifically, we considered two samples with 10032 and 47473 atoms, respectively. Heat correlations were computed along six independent MD runs as long as 100ns to get converged conductivities.
Our calculated values are 2500 $\pm$ 120 W/(m$\cdot$K) and 2660 $\pm$ 130 W/(m$\cdot$K), for the smaller and larger system, respectively. 
As a matter of fact, they are in really good agreement with previous estimations, which range in the interval 2600-3050 W/(m$\cdot$K) \cite{xu2014length, MORTAZAVI20151, Bagri, MORTAZAVI2016318, PhysRevB.95.144309, doi:10.1021/nn200114p, doi:10.1021/nl0731872, ghosh2010dimensional, PhysRevB.92.094301, PhysRevB.81.205441, doi:10.1021/nl102923q} and have been obtained with a similar computational setup, namely: by equilibrium MD and optimized Tersoff potential.  
Interesting enough, a similar good agreement is also found with other investigations where a different simulation protocol was used: by means of non-equilibrium MD simulations, a value for thermal conductivity of pristine graphene was obtained in the range 2600- 3050 W/(m$\cdot$K)) \cite{xu2014length, MORTAZAVI20151, Bagri}.
All simulation data lie well within the experimentally reported range  of $\sim$2500 - 5000 W/(m$\cdot$K) \cite{ghosh2010dimensional, doi:10.1021/nl9041966, Seol213, doi:10.1021/nl102923q}.

\begin{acknowledgement}
The authors acknowledge the project: ModElling Charge and Heat trANsport in 2D-materIals based Composites - MECHANIC
reference number: PCI2018-093120 funded by Ministerio de Ciencia, Innovaci\'on y Universidades.
ICN2 is funded by the CERCA Programme/Generalitat de Catalunya and is supported by the Severo Ochoa program from Spanish MINECO (Grant No. SEV-2017-0706).
\end{acknowledgement}



\bibliographystyle{achemso}
\bibliography{biblio}

\end{document}